\documentclass{article}
\usepackage{spconf,amsmath,graphicx}
\usepackage{acronym}
\usepackage{svg}
\usepackage{cite}
\usepackage{censor} 
\usepackage{flushend} % distribute references in 2 cols
\usepackage[citecolor=black,%black!40!red,
            colorlinks=true,
            linkcolor=black,%blue!10!black,
            urlcolor  = black,
            pdfpagelabels={true},
            pdftitle={Non-Intrusive Speech Intelligibility Prediction for Hearing Impaired Individuals using Self-Supervised Speech Representations},
            pdfauthor={George Close, Thomas Hain, Stefan Goetze},
            pdfkeywords={hearing loss, metric prediction, neural networks, self-supervised speech representations},
            pdfsubject={hearing loss, metric prediction, neural networks, self-supervised speech representations}]
            {hyperref}
\usepackage{tabu}
\usepackage[english]{babel}
\addto\extrasenglish{}
\addto\extrasenglish{}
\addto\extrasenglish{}
\addto\extrasenglish{}
\addto\extrasenglish{}
\addto\extrasenglish{}
\addto\extrasenglish{}
\acrodef{ASR}  {automatic speech recognition}
\acrodef{BLSTM}{bidirectional long short-term memory}
\acrodef{CEC}  {Clarity Enhancement Challenge}
\acrodef{CNN}  {convolutional neural network}
\acrodef{CPC}  {Clarity Prediction Challenge}
\acrodef{CPC1} {Clarity Prediction Challenge 1}
\acrodef{dBHL} {decibels hearing level}
\acrodef{DFT}  {discrete Fourier transform}
\acrodef{GAN}  {Generative Adversarial Network}
\acrodef{h2h}  {human-to-human}
\acrodef{h2m}  {human-to-machine}
\acrodef{HA}   {hearing aid}
\acrodef{HL}   {hearing loss}
\acrodef{HLS}  {hearing loss simulator}
\acrodef{HASPI}{Hearing-Aid Speech Perception Index}
\acrodef{HSR}  {human speech recognition}
\acrodef{MBSTOI}{Modified Binaural Short-Time Objective Intelligibility}
\acrodef{MOS}  {mean opinion score}
\acrodef{MMSE} {minimum mean squared error}
\acrodef{MSE}  {mean squared error}
\acrodef{NN}   {neural network}
\acrodef{OLA}  {Overlap-Add}
\acrodef{PESQ} {Perceptual Evaluation of Speech Quality}
\acrodef{RMSE} {Root Mean Square Error}
\acrodef{SE}   {speech enhancement}
\acrodef{SI}   {speech intelligibility}
\acrodef{SNR}  {Signal to Noise Ratio}
\acrodef{SPIN} {Speech In Noise}
\acrodef{SSSR} {self-supervised speech representation}
\acrodef{STFT} {short time Fourier transform}
\acrodef{STOI} {Short-Time Objective Intelligibility}
\acrodef{VAD}  {Voice Activity Detector}
\acrodef{WER}  {Word Error Rate}
\acrodef{XLSR} {Cross-Lingual Speech Representation}
\title{Non-Intrusive Speech Intelligibility Prediction for Hearing Impaired Individuals using Self-Supervised Speech Representations\thanks{This work was supported by the Centre for Doctoral Training in Speech and Language Technologies (SLT) and their Applications funded by UK Research and Innovation [grant number EP/S023062/1]. This work was also funded in part by TOSHIBA Cambridge Research Laboratory.}}
\name{George Close, Thomas Hain, and Stefan Goetze}
\address{Speech and Hearing Group, Department of Computer Science, University of Sheffield, United Kingdom}
\begin{document}
%\ninept
%
\maketitle
\begin{abstract}
% Self-supervised speech representations are recently successfully applied to a number of speech-processing tasks. In particular, they have been found to be useful as a feature extractor for the prediction of human speech quality evaluations, which is in turn  highly relevant for assessment and training of speech enhancement systems for users with normal or impaired hearing. However, the exact knowledge of why and how quality-related information is encoded well in these representations remains poorly understood. In this work, the task of non-intrusive prediction of labels obtained from listening tests from assessors with normal hearing is extended to the prediction of intelligibility ratings for hearing-impaired individuals using data from the  Clarity Prediction Challenge $1$. It is found that self-supervised speech representations are useful as input features to non-intrusive hearing loss models, outperforming spectrogram representations  achieving competitive performance to significantly more complex systems. However, questions remain about their adaptation to unseen data and their ability to capture listener-specific hearing loss information. 
Self-supervised speech representations (SSSRs) have been successfully applied to a number of speech-processing tasks, e.g. as feature extractor for speech quality (SQ) prediction, which is, in turn, relevant for assessment and training speech enhancement systems for users with normal or impaired hearing. However, exact knowledge of why and how quality-related information is encoded well in such representations remains poorly understood. In this work, techniques for non-intrusive prediction of SQ ratings are extended to the prediction of intelligibility for hearing-impaired users. It is found that self-supervised representations are useful as input features to non-intrusive prediction models, achieving competitive performance to more complex systems. A detailed analysis of the performance depending on Clarity Prediction Challenge 1 listeners and enhancement systems indicates that more data might be needed to allow generalisation to unknown systems and (hearing-impaired) individuals.%However, questions remain about their adaptation to unseen data.
\end{abstract}

\section{Introduction}
Age-related \ac{HL} is an increasingly prevalent problem in countries with ageing populations worldwide.
In the United Kingdom, for example, approximately $12$ million people suffer from \ac{HL} of greater than 25 \ac{dBHL}; by $2035$ this is predicted to increase to $14.2$ million \cite{park2020population}. \Acl{HL} can often impede an individual's ability to participate in a spoken conversation, especially in noisy environments, as parts of the speech can become unintelligible~\cite{Voelker_SI_2015}.  As such, the development of methods to increase \ac{SI} in assistive listening devices to alleviate this is of paramount importance~\cite{GXRRA10}. While there have been large improvements in speech enhancement technology thanks to neural network-based approaches \cite{dubey2023icassp,fu2021metricganu,gc-mg-minus} these can often be challenging to implement in small form factor \ac{HA} hardware. Furthermore, given that the exact severity and nature of hearing loss differs greatly between individuals, a 'one size fits all' approach is not viable.  
The Clarity Project \cite{clarity_project} aims to improve the design of hearing aids via two alternating challenges and related datasets \cite{clarity_challenge}, the \ac{CEC} and the \ac{CPC}. The \ac{CEC} is concerned with the design of the actual enhancement algorithm, while the \ac{CPC} involves the prediction of the intelligibility of the enhanced speech for hearing-impaired listeners. The overall aim of the \ac{CPC} is to produce systems that mimic the behaviour of hearing-impaired listeners, reducing the need for expensive and time-consuming human assessment by listening tests, while also providing a training target or metric for enhancement systems.\\
\Acp{SSSR} have been found to be useful either as pretrained layers or feature transformations in many speech-related applications \cite{wav2vec,hubert,superb}. It is understood that \ac{SSSR}s are able to encode and predict the context of the speech content in the input audio, and thus model the patterns of spoken language. Recent work \cite{gc-sssr_loss, nisqa_pretrained_ss, perceptual_quality_phone_fort,10097097,pasad2023comparative} has found that in addition to speech content, \ac{SSSR}s are also able to capture information on potentially corrupting noise and distortion in the input audio. \\
In this work, \acp{SSSR} are used as a feature transformation for non-intrusive neural speech intelligibility prediction networks, trained on the CPC1 challenge dataset. Non-intrusive metric prediction networks using \acp{SSSR} are proposed to serve as feature extractors and analysed for different latent or output \ac{SSSR} layers to predict \ac{SI} for hearing-impaired users.\\
The remainder of the paper is structured as follows. In \autoref{sec:sssrs}, the \acp{SSSR} used in this work are briefly introduced and \autoref{sec:metric_prediction} reviews the use of \acp{SSSR} in related tasks. \autoref{sec:dataset} introduces the dataset as used in this work and in \autoref{sec:relationsips} the relationships between this dataset and \ac{SSSR} distance measures are examined.  Finally, in Sections~\ref{sec:experiments} and \ref{sec:results}, experiments involving the use of \ac{SSSR} as feature representations in non-intrusive intelligibility prediction networks are described, and the results analysed.

\section{Self Supervised Speech Representations}\label{sec:sssrs}

\begin{figure}[!ht]
 \centering
 \resizebox{1\columnwidth}{!}{%
 \graphicspath{{figs}} 
 %\resizebox{\columnwidth}{!}{% this can be used for final
 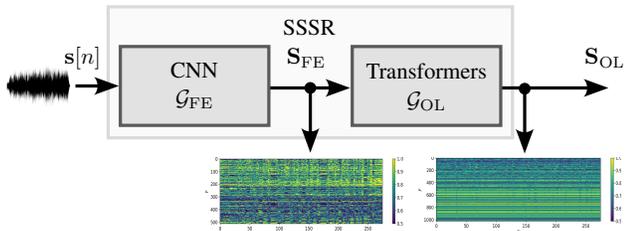 
 }
\caption{Representations extracted from \ac{SSSR} model with time-domain input signal $\mathbf{s}[n]$. Feature channels are sorted \cite{atttasnet} and values normalised for clarity.}
\label{fig:sssr_example}
\end{figure}

Generally, \acp{SSSR} are neural networks that, given a waveform representation of speech audio $\mathbf{s}[n]$, produce a final output which expresses the \emph{context} of that input speech audio. As the name suggests, they are trained in a self-supervised way, typically by \emph{masking} some segment of the input audio representation and tasking the network in training to recreate the masked segment \cite{wav2vec}. Structurally, the networks consist of two distinct stages as shown in \autoref{fig:sssr_example}. 

The input waveform $\mathbf{s}[n]$ with discrete time index $n$ is first processed by a number of 1-D convolutional layers, resulting in a two-dimensional feature encoding representation. In the second stage, this representation is processed by a number of Transformer~\cite{transformer} layers, to give the final two-dimensional output. In both stages, one of the dimensions represents time, while the other represents features. 
For a time domain signal $\mathbf{s}[n]$, the output of the initial \ac{CNN} encoder stage $\mathcal{G}_\mathrm{FE}$  of a \ac{SSSR} is
\begin{equation}
    \label{eq:fe_transform}
    \mathbf{S}_\mathrm{FE} = \mathcal{G}_\mathrm{FE}(\mathbf{s}[n]),
\end{equation} 
where operator $\mathcal{G}_\mathrm{FE}$ denotes the 1-D convolutional layers encompassing the encoder.
The subsequent processing by the Transformer based stage can be defined by an additional operator $\mathcal{G}_\mathrm{OL}$ denoting the Transformer layers encompassing the final output stage, i.e.
\begin{equation}
    \label{eq:ol_transform}
    \mathbf{S}_\mathrm{OL} = \mathcal{G}_\mathrm{OL}(\mathcal{G}_\mathrm{FE}(\mathbf{s}[n])).
\end{equation}
Both signal representations $\mathbf{S}_\mathrm{FE}$ and $\mathbf{S}_\mathrm{OL}$ have two dimensions: time $T$, depending on the length of the input audio in block time, and feature dimension $F$.\\

%\subsection{Cross-Lingual Speech Representation (XLSR)}
The \emph{\ac{XLSR}}~\cite{xlsr} is one of the \ac{SSSR} representations used in this work. It is a variant of the Wav2Vec2~\cite{wav2vec} structure. It is trained on $436,000$ hours of audio data from a number of languages, including the BABEL dataset which contains potentially noisy telephone conversation recordings. Its network is structured in the way described above, with the outputs of $\mathcal{G}_\mathrm{FE}$ having a feature dimension of $512$ and the final (transformer) outputs of $\mathcal{G}_\mathrm{OL}$ having a feature dimension of $1024$. In this work, the smallest version of XLSR sourced from the HuggingFace\footnote{\url{https://huggingface.co/facebook/wav2vec2-xls-r-300m}} is used.\\
%\subsection{Hidden Unit BERT (HuBERT)}
\emph{Hidden Unit Bidirectional Encoder Representations from Transformers (HuBERT)} \cite{hubert} differ w.r.t.~the general training of a \ac{SSSR} described above in that the training target is a cluster of masked frames similar to BERT~\cite{bert} rather than the masked frame itself. However, its network structure follows the same pattern, with the outputs of $\mathcal{G}_\mathrm{FE}$ having a feature dimension of $512$ and the final output representation a feature dimension of $768$. In this work, we use the HuBERT Large model trained on $960$ hours of clean English speech sourced from the LibriSpeech~\cite{7178964} dataset, from the Fairseq GitHub repository\footnote{\url{https://github.com/facebookresearch/fairseq}}. 

\section{SSSRs for Metric Prediction}\label{sec:metric_prediction}
SSSRs have been applied to metric prediction tasks, typically to quality prediction \cite{NISQA,sssr_quality1}.
In \cite{nisqa_pretrained_ss}, XLSR representations are used as feature extraction in a non-intrusive human MOS prediction network. 
%The proposed system performs very well on the ConferencingSpeech 2022 quality prediction challenge~\cite{yi22b_interspeech}.
Similarly, in \cite{becerra22_interspeech} \ac{SSSR}s are used for the same quality prediction task, but they are fine-tuned with a mean pooling layer rather than being used simply as feature extraction. \acp{SSSR} were also applied to the CPC1 challenge in \cite{edozezario22_interspeech}, where they were used as feature extractors alongside spectrograms and learnable filter banks. \\
In all these cases, only the final \ac{SSSR} output $\mathcal{G}_\mathrm{OL}$ was considered. However, findings in \cite{gc-sssr_loss} suggest that the output of the initial encoding stage $\mathcal{G}_\mathrm{FE}$ better captures quality-related information. As such, in this work, both representations stages are considered and compared as feature transformations for speech intelligibility prediction. 
\section{Clarity Challenge 1 Data}\label{sec:dataset}
\begin{figure}[!ht]
 \centering
 %\resizebox{0.8\columnwidth}{!}{%
 %\def\svgwidth{\columnwidth} 
 \graphicspath{{figs}} 
 %\resizebox{\columnwidth}{!}{% this can be used for final
 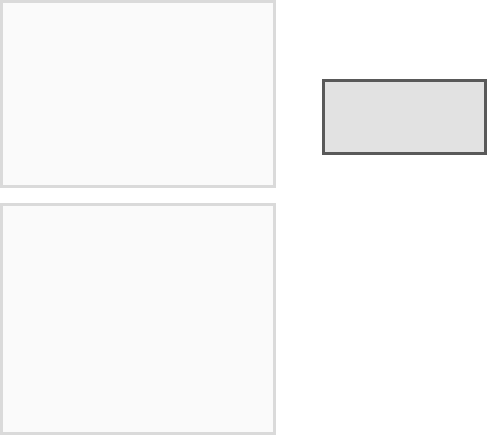 

\caption{Signal generation for \acl{CPC}.\label{fig:Clarity_schematic}}
\end{figure}

\begin{figure}[!ht]
 \centering
 %\resizebox{0.8\columnwidth}{!}{%
 \def\svgwidth{\columnwidth} 
 \graphicspath{{figs}} 
 %\resizebox{\columnwidth}{!}{% this can be used for final
 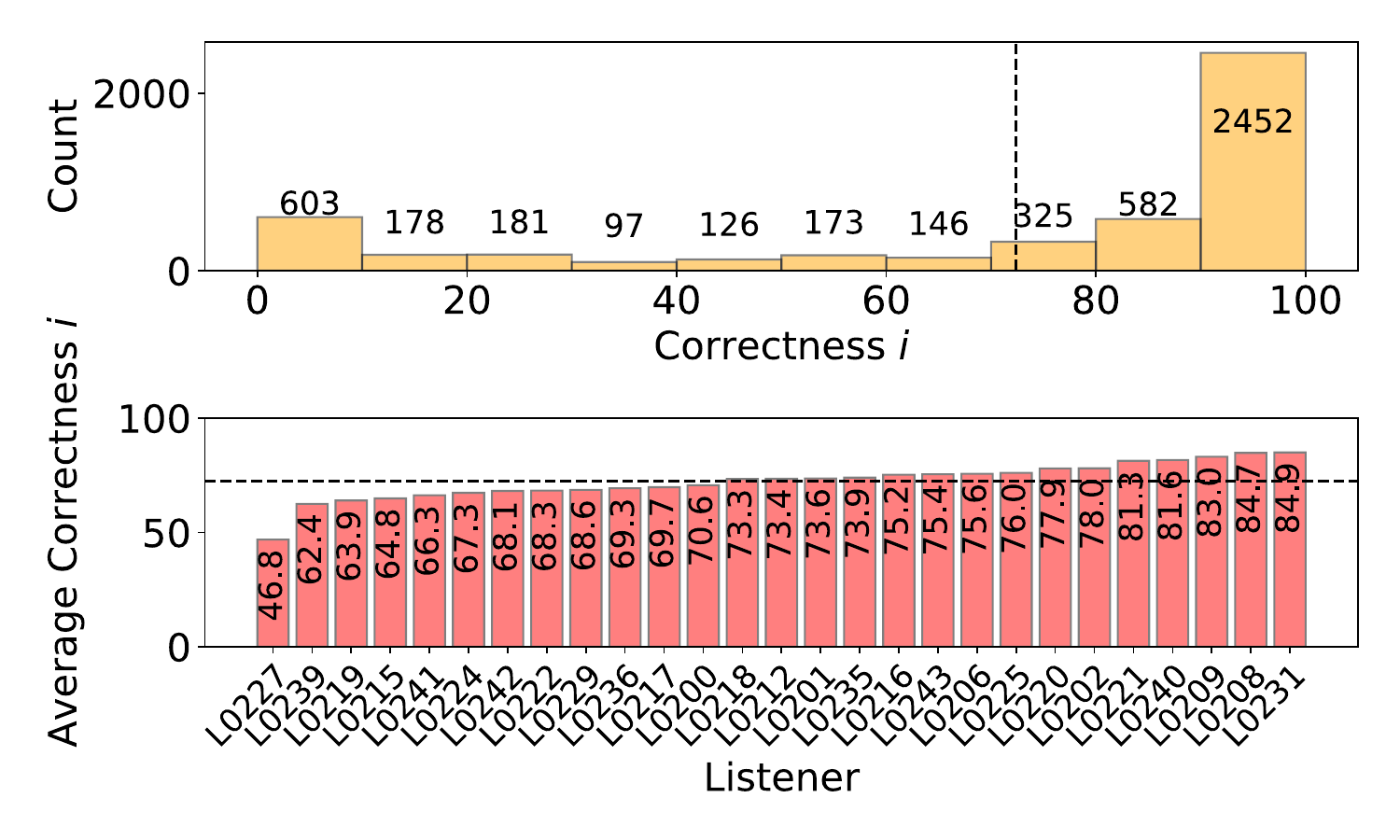 

\caption{Histogram showing the distribution of ground truth correctness $i$ in CPC1 training set (top) and a bar chart showing average correctness $i$ per listener in the CPC1 training set (bottom). Dotted lines are respective overall average values.}\label{fig:intel_bar}
\end{figure}

The dataset for the first \ac{CPC1}\cite{barker22_interspeech} as used in this work can be expressed as a series of sequences: 
%tuples:
($\mathbf{\hat{s}}[n]$, $\{\mathbf{a}_l,\mathbf{a}_r\}$, $i$), which is generated as visualised in \autoref{fig:Clarity_schematic}.
$\mathbf{\hat{s}}[n]$ represents the output of a hearing aid system for some noisy speech input $\mathbf{x}[n]$, containing some clean speech $\mathbf{s}[n]$.
$\{\mathbf{a}_l,\mathbf{a}_r\}$ are the left and right audiogram representations of a particular listener's hearing loss. Blue and red box plots in \autoref{fig:Clarity_schematic} illustrate the \ac{HL} distribution in the \ac{CPC1} dataset from which the individual audiograms are sampled. Finally, $i$ represents the intelligibility of the audio $\mathbf{\hat{s}}[n]$ for that listener, defined as the percentage of words they were able to reproduce by speaking aloud immediately after hearing the audio, compared to a ground truth transcription of the speech also denoted as the \emph{correctness} of the listener's response. Additionally, audio $\mathbf{\hat{s}}'[n]$ is defined as the output of the baseline Cambridge MSBG \ac{HLS}, denoted here by operator $\mathcal{S}$, cf.~\cite{Stone1999TolerableHA} for additional details on the clarity system.
\begin{equation}
\label{eq:hl_sim}
\mathbf{\hat{s}}'[n] = \mathcal{S}(\mathbf{\hat{s}}[n], \{\mathbf{a}_l,\mathbf{a}_r\})
\end{equation}
The signal $\mathbf{\hat{s}}'[n]$ is an approximation of the audio that is perceived by the hearing-impaired listener. This can be thought of as encoding the hearing characteristics of the specific listener (audiogram) within the signal. Note that all signals in the dataset are binaural i.e.~consist of left and right channels, denoted by $l$ and $r$, respectively. \\
The upper plot in \autoref{fig:intel_bar} shows the distribution of correctness $i$ in the CPC1 training set. 
From this, it can be observed that in the majority of cases, the listener was able to fully reproduce the speech in the audio they heard, i.e.~$i = 100$ for $\approx 50 \%$ of the assessed files. The next largest class is where $i = 0$, meaning that the listener was not able to reproduce any words in the audio. This distribution is due to the more realistic \textit{in-the-wild} \ac{SI} measurement strategy for the Clarity dataset \cite{barker22_interspeech} which is in contrast to lab-based \ac{SI} matrix tests \cite{SI_MatrixTests_Multilingual}. The lower panel of \autoref{fig:intel_bar} shows the average correctness $i$ for each listener in the CPC1 training set. With the exception of listener L0227, all of the listeners achieve similar performance. \\
% \autoref{fig:audiogram} shows the audiograms for the listeners in the CPC1 data, in relation to the frequency characteristics of speech phones, with the grey shaded area depicting the typical level range of these phones. 

% \begin{figure}[!ht]
%  \centering
%  %\resizebox{0.8\columnwidth}{!}{%
%  \def\svgwidth{\columnwidth} 
%  \graphicspath{{figs}} 
%  %\resizebox{\columnwidth}{!}{% this can be used for final
%  \input{figs/corr_hist.pdf_tex} 

% \caption{Histogram showing distribution of ground truth correctness $i$ in CPC1 Training set}\label{fig:intel_hist}
% \end{figure}
% \begin{figure}[!ht]
%  \centering
%  \resizebox{1.1\columnwidth}{!}{%
%  \def\svgwidth{\columnwidth} 
%  	\graphicspath{{figs/}} 
%  \input{figs/audiogram2.pdf_tex} 
%  }
% \caption{
% {Audiogram data for} 
% \textcolor{red}{left (red)} {and} 
% \textcolor{blue}{right (blue)} {ear of}  CPC1 listener{s}  
% and the frequency characteristics of speech phones 
% } \label{fig:audiogram}
% \end{figure}

\section{Analysing Relationships between SSSRs and Human Speech Intelligibility}\label{sec:relationsips}
In order to express the relationship between \acp{SSSR} and correctness $i$ in the dataset, two distance measures are defined in a \ac{MSE} sense:
\begin{equation}
\label{eq:fe_mse}
    d_{\mathrm{FE}}  =\frac{1}{T F} \sum_{t}^{T} \sum_{f}^{F} (\mathbf{S}_{\mathrm{FE}}[t,f] - \mathbf{{P}}_{\mathrm{FE}}[t,f])^2 
\end{equation}
\begin{equation}
\label{eq:ol_mse}
    d_{\mathrm{OL}}  =\frac{1}{T F} \sum_{t}^{T} \sum_{f}^{F} (\mathbf{S}_{\mathrm{OL}}[t,f] - \mathbf{{P}}_{\mathrm{OL}}[t,f])^2 
\end{equation}
The distance $d_{\mathrm{FE}}$ in (\ref{eq:fe_mse}) expresses the \ac{MSE} distance between the \ac{SSSR} \textit{feature encoding} layer representations $\mathbf{S}_{\mathrm{FE}}[t,f]$  of the clean reference audio $\mathbf{s}[n]$ and the representations $\mathbf{{P}}_{\mathrm{FE}}[t,f]$ of the test signal $\mathbf{p}[n]$, while (\ref{eq:ol_mse}) expresses the \ac{MSE} distance between the \ac{SSSR} \textit{output layer} representations $\mathbf{S}_{\mathrm{OL}}[t,f]$ and $\mathbf{{P}}_{\mathrm{OL}}[t,f]$, with $t$ and $f$ denoting block time and feature index, respectively. Note that $\mathbf{p}[n]$ and is a placeholder for either the speech signal after \ac{HA} enhancement $\mathbf{\hat{s}}[n]$ or this signal after \ac{HLS} processing $\mathbf{\hat{s}}'[n]$ as shown in \autoref{fig:Clarity_schematic}. 
Distances in (\ref{eq:fe_mse}), (\ref{eq:ol_mse}) are designed to express the distortion captured by the \ac{SSSR} due to the transformations which have been applied to $\mathbf{s}[n]$ to produce e.g.~$\mathbf{\hat{s}}'[n]$, i.e.~the artificial distortion/reverb added to create $\mathbf{x}[n]$, enhancement by the hearing aid system (in $\mathbf{\hat{s}}[n]$) and finally the \ac{HLS}. In addition to distances (\ref{eq:fe_mse}) and (\ref{eq:ol_mse}) the \ac{MSE} distance between spectrogram representations of $\mathbf{s}[n]$ and $\mathbf{p}[n]$,
\begin{equation}
    \label{eq:spec_mse}
    d_{\mathrm{SG}} = 
    \frac{1}{T F_{\mathrm{Hz}}} 
    \sum_{t}^{T} 
    \sum_{f_{\mathrm{Hz}}}^{F_{\mathrm{Hz}}} 
    \left(\mathbf{S}_{\mathrm{SG}}[t,f_{\mathrm{Hz}}] - \mathbf{P}_{\mathrm{SG}}[t,f_{\mathrm{Hz}}]\right)^2, 
\end{equation}
will be analysed, with $f_{\mathrm{Hz}}$ and $F_{\mathrm{Hz}}$ denoting the technical frequency and the highest frequency analysed, respectively. In the following, the left (first) channel of the audio is used to compute the distance measures (\ref{eq:fe_mse}), (\ref{eq:ol_mse}) and (\ref{eq:spec_mse}). 
% \begin{table}[!ht]
% \caption{Spearman and Pearson correlations between  $(\mathbf{s}[n]$,$ \mathbf{\hat{s}}[n])$, $(\mathbf{s}[n]$,$ \mathbf{\hat{s}}'[n])$  distance measures and correctness values $i$ in the CPC1 training set}
% \centering
% \begin{tabular}{|l|c|c|}
% \hline
% Measure           & Spearman & Pearson \\ \hline
% $\mathrm{SPEC}_\mathrm{MSE}(\mathbf{s}[n], \mathbf{\hat{s}}[n])$      & -0.10    & -0.18   \\ 
% $\mathrm{SPEC}_\mathrm{MSE}(\mathbf{s}[n], \mathbf{\hat{s}}'[n])$      & -0.09    & -0.07   \\ \hline
% XLSR $\mathrm{FE}_\mathrm{MSE}(\mathbf{s}[n], \mathbf{\hat{s}}[n])$   & -0.13    & -0.16   \\ 
% XLSR $\mathrm{FE}_\mathrm{MSE}(\mathbf{s}[n], \mathbf{\hat{s}}'[n])$   & -0.24    & -0.28   \\ 
% XLSR $\mathrm{OL}_\mathrm{MSE}(\mathbf{s}[n], \mathbf{\hat{s}}[n])$   & -0.26    & -0.27   \\
% XLSR $\mathrm{OL}_\mathrm{MSE}(\mathbf{s}[n], \mathbf{\hat{s}}'[n])$   & -0.24    & -0.24   \\ \hline
% HuBERT $\mathrm{FE}_\mathrm{MSE}(\mathbf{s}[n], \mathbf{\hat{s}}[n])$ & \textbf{-0.38}    & \textbf{-0.47}   \\ 
% HuBERT $\mathrm{FE}_\mathrm{MSE}(\mathbf{s}[n], \mathbf{\hat{s}}'[n])$ & -0.23    & -0.29   \\ 
% HuBERT $\mathrm{OL}_\mathrm{MSE}(\mathbf{s}[n], \mathbf{\hat{s}}[n])$ & {-0.10}    & {-0.17}   \\
% HuBERT $\mathrm{OL}_\mathrm{MSE}(\mathbf{s}[n], \mathbf{\hat{s}}'[n])$ & {-0.28}    & {-0.32}   \\ \hline
% \end{tabular}
% \end{table}
\begin{table}[!ht]
\caption{Spearman and Pearson correlations between 
%$(\mathbf{s}[n]$,$ \mathbf{\hat{s}}[n])$, $(\mathbf{s}[n]$,$ \mathbf{\hat{s}}'[n])$
distance measures and correctness values $i$ in the CPC1 training set, strongest correlations in bold.}\label{tab:mse_corrs}
\centering
\resizebox{1\columnwidth}{!}{%
\begin{tabular}{l|c|c|c}
%\hline
\textbf{Representation, Distance}          & $\mathbf{p}[n]$ &\textbf{Spearman} & \textbf{Pearson} \\ \hline
SPEC, $d_\mathrm{SG}$, (\ref{eq:spec_mse}) & $\mathbf{\hat{s}}[n]$& $-0.10$    & $-0.18$   \\ 
SPEC, $d_\mathrm{SG}$, (\ref{eq:spec_mse}) & $\mathbf{\hat{s}}'[n]$     & $-0.09$    & $-0.07$   \\ \hline
$\mathrm{XLSR}$, $d_\mathrm{FE}$, (\ref{eq:fe_mse}) & $\mathbf{\hat{s}}[n]$&  $-0.13$    & $-0.16$   \\ 
$\mathrm{XLSR}$, $d_\mathrm{FE}$, (\ref{eq:fe_mse})& $\mathbf{\hat{s}}'[n]$& $-0.24$    & $-0.28$   \\ 
$\mathrm{XLSR}$, $d_\mathrm{OL}$, (\ref{eq:ol_mse}) & $\mathbf{\hat{s}}[n]$& $-0.26$    & $-0.27$   \\
$\mathrm{XLSR}$, $d_\mathrm{OL}$, (\ref{eq:ol_mse})   & $\mathbf{\hat{s}}'[n]$& $-0.24$    & $-0.24$   \\ \hline
$\mathrm{HuBERT}$, $d_\mathrm{FE}$, (\ref{eq:fe_mse}) & $\mathbf{\hat{s}}[n]$& $\mathbf{-0.38}$    & $\mathbf{-0.47}$   \\ 
$\mathrm{HuBERT}$, $d_\mathrm{FE}$, (\ref{eq:fe_mse}) & $\mathbf{\hat{s}}'[n]$& $-0.23$    & $-0.29$   \\ 
$\mathrm{HuBERT}$, $d_\mathrm{OL}$, (\ref{eq:ol_mse}) & $\mathbf{\hat{s}}[n]$& ${-0.10}$    & ${-0.17}$   \\
$\mathrm{HuBERT}$, $d_\mathrm{OL}$, (\ref{eq:ol_mse}) & $\mathbf{\hat{s}}'[n]$& ${-0.28}$    & ${-0.32}$   \\ %\hline
\end{tabular}
}
\end{table}

\noindent \autoref{tab:mse_corrs} shows the Spearman and Pearson correlations of the MSE distances (\ref{eq:fe_mse})-(\ref{eq:spec_mse}) with the correctness values $i$ for the CPC1 training set. Absolute correlations are low, but this is expected for the Clarity dataset (cf.~\cite{barker22_interspeech} and \autoref{sec:dataset}). Comparing the distances between the feature representations in (\ref{eq:fe_mse})-(\ref{eq:spec_mse}) and the intelligibility scores $i$  allows for an expression of how distortion in the signal, which might affect intelligibility, is captured by that feature representation. Interestingly, applying the hearing loss simulation $\mathcal{S}$ in (\ref{eq:hl_sim}) does not uniformly improve the correlation with $i$ across all distances in \autoref{tab:mse_corrs}; only for the XLSR encoder output representation distance $d_\mathrm{FE}$ and the HuBERT final output representation distance $d_\mathrm{OL}$ does using $\mathbf{\hat{s}}'[n]$ lead to higher correlation than using $\mathbf{\hat{s}}[n]$. 
\section{SSSR-based Intelligibility Prediction}\label{sec:experiments}
This section proposes the use of \ac{SSSR}s as features in non-intrusive neural intelligibility prediction networks. Following the findings from \autoref{tab:mse_corrs}, both the hearing aid output signal $\mathbf{\hat{s}}[n]$ and that signal processed by the hearing loss simulation $\mathbf{\hat{s}}'[n]$ are used as the input audio to the models, as no conclusive best representation is indicated by these results.
\subsection{Dataset}
Models are trained on both the open and closed training and test sets detailed in the CPC1 description~\cite{clarity_challenge}. The closed set has the same listeners and systems for both train and testsets, while the open set has $5$ unseen listeners and $1$ unseen system in its testset. For more detail as to how these sets are differentiated, see \cite{clarity_challenge}. 
A validation set is created using $10\%$ of the available training data. As we are interested in non-intrusive predictors, either the hearing aid output signal $\mathbf{\hat{s}}[n]$ or the hearing loss simulated audio signal $\mathbf{\hat{s}}'[n]$ are used to predict the Correctness label $i$. For the closed set, the training set contains $4376$ utterances, the validation set $487$ and the test set $2421$. The training set contains $3222$ utterances for the open set, $358$ for the validation set, and $632$ for the test set. 

\subsection{Model Structure and Experiment Setup}
A model structure inspired by \cite{nisqa_pretrained_ss} is chosen for the \ac{SI} prediction network. Five feature extraction methods are used; outputs of $\mathcal{G}_\mathrm{FE}$ and $\mathcal{G}_\mathrm{OL}$ for both, HuBERT and XLSR representations, as well as a spectrogram representation denoted as $\mathrm{SPEC}$. After the feature extraction, the resultant representation is processed by $2$ \ac{BLSTM} layers with an input size equal to the feature dimension $F$  of the input and a hidden layer size of $F/2$. The final layer is an attention pooling feed-forward layer, similar to that in NISQA~\cite{NISQA} with a single output neuron and a sigmoid activation to output the predicted correctness  $\hat{i}$ (normalised between $0$ and $1$). Note that due to different dimensions $F$ of different feature representations, the number of parameters in each network varies from $923,906$ for the models using spectrogram representations to as many as $14,701,570$ for the models using the XLSR output layer, i.e.~$\mathcal{G}_\mathrm{OL}$.\\
The two input audio representations $\mathbf{\hat{s}}[n]$ or $\mathbf{\hat{s}}'[n]$ are used, i.e.~the output of the hearing aid systems and the enhanced audio processed by the hearing loss simulation, as in (\ref{eq:hl_sim}). 
As these audio representations have two channels, each channel is processed by the model separately; during training, the loss for each channel is computed and then summed before being back-propagated to the model. During validation and testing, the maximum value between each channel is taken as an approximation of the \emph{better ear effect}~\cite{zurek1993binaural}.\\ 
The spectrogram representation is created by a \ac{STFT} with a window length of $20$~ms, a hop length of $10$~ms and an FFT size of $1024$. All audio is re-sampled 
%from the original 44.1kHz
to $16$~kHz such that it can be used as inputs to the \ac{SSSR} models. \\ 
\section{Results}\label{sec:results}
In addition to the intrusive (reference-signal-based) challenge baseline, the best-performing non-intrusive entries to the challenge are reported in this section as additional baselines, as the proposed system is also non intrusive. Challenge entry E23~\cite{McKinney_2022} makes use of contrastive predictive coding and vector quantisation features. E06~\cite{close22_interspeech} is similar to the proposed system, denoted by $\mathrm{SPEC}$ in the following, but uses a CNN based network structure. E33~\cite{edozezario22_interspeech} also utilises \acp{SSSR} as feature extraction, but spectrogram and learnable filterbank features are also used as model inputs. E29~\cite{tu22b_interspeech} makes use of an information-theory-inspired approach, wherein the difference between internal representations in neural \ac{ASR} systems is used to approximate human intelligibility, and was the overall best non-intrusive challenge entry. 
\subsection{Results on CPC1 Closed set}
\begin{table}[]
\caption{Non-Intrusive Prediction Performance on the \ac{CPC1} closed set. Best performances for baselines and proposed methods in boldface font.}
\label{tab:closed-results}
\begin{tabular}{l|cccc}
\textbf{Model Name} & \multicolumn{1}{l}{\textbf{RMSE}} & \multicolumn{1}{l}{\textbf{Var}} & \multicolumn{1}{l}{\textbf{Spearman}} & \multicolumn{1}{l}{\textbf{Pearson}} \\ \hline
\textit{CPC1 Baseline}            & \textit{28.50}                             &            \textit{--}                      & \textit{0.62}                                  &    \textit{--}                                  \\
\textit{E23}~\cite{McKinney_2022}                 & \textit{41.50}                             & \textit{--}                               & \textit{0.07}                                  & \textit{--}                                   \\
\textit{E06}~\cite{close22_interspeech}                 & \textit{32.00}                             & \textit{--}                               & \textit{0.43}                                  & \textit{--}                                 \\
\textit{E33}~\cite{edozezario22_interspeech}                  & \textit{24.10}                            & \textit{--}                               & \textit{0.75}                                  & \textit{--}                                   \\ 
\textit{E29}~\cite{tu22b_interspeech}       & \textit{\textbf{23.30}}           & \textit{--}                      & \textit{\textbf{0.77}}                & \textit{--}                          \\\hline
$\mathrm{SPEC}$ ~ $\mathbf{\hat{S}}_\mathrm{SPEC}$ & 25.45	 & 0.52 &	0.59 &	0.72                               \\
$\mathrm{SPEC}$ ~$\mathbf{\hat{S}'}_\mathrm{SPEC}$               & 25.45                             & 0.52                             & 0.58                                  & 0.72                                 \\
$\mathrm{HuBERT}~\mathbf{\hat{S}}_\mathrm{FE}$      &30.82&	0.61&	0.44&	0.56                                 \\
$\mathrm{HuBERT}~\mathbf{\hat{S}'}_\mathrm{FE}$       & 26.64                             & 0.53                             & 0.56                                  & 0.70                                 \\
$\mathrm{HuBERT}~\mathbf{\hat{S}}_\mathrm{OL}$     &    \textbf{24.76}&	\textbf{0.50}&	0.59 &	\textbf{0.74}\\
$\mathrm{HuBERT}~ \mathbf{\hat{S}'}_\mathrm{OL}$         & 24.82                             & \textbf{0.50}                             & \textbf{0.61}                                  & \textbf{0.74}                                 \\
$\mathrm{XLSR}~ \mathbf{\hat{S}}_\mathrm{FE}$     &   25.01&	\textbf{0.50}&	0.60 &	\textbf{0.74}                        \\
$\mathrm{XLSR}~ \mathbf{\hat{S}'}_\mathrm{FE}$         & 25.33                             & 0.51                             & 0.60                                  & 0.72                                 \\
$\mathrm{XLSR}~ \mathbf{\hat{S}}_\mathrm{OL}$   &   28.42	&0.58&	0.47&	0.66 \\
$\mathrm{XLSR}~ \mathbf{\hat{S}'}_\mathrm{OL}$      &   30.20 &	0.61& 	0.52&	0.64
\end{tabular}
\end{table}
\autoref{tab:closed-results} shows the performance of the proposed systems for the CPC1 closed set. All proposed systems show comparable performance with the best-performing challenge entries, although, none of the proposed systems outperforms system E29. It should be noted, however, that the computation overhead to implement system E29 is significantly greater than any of the proposed systems here, as several state-of-the-art \ac{ASR} systems must be trained and fine-tuned for E29. Of the proposed systems trained on the outputs of the hearing loss simulation $\mathbf{\hat{s}}'[n]$, the best performing is the model which uses HuBERT output representations $\mathbf{\hat{S}'}_\mathrm{OL}$ as features. This is consistent with the findings in \autoref{tab:mse_corrs} which shows that the distance measure using this representation had the highest correlation with $i$ of those distances computed using $\mathbf{\hat{s}}'[n]$. Of those trained using the hearing aid outputs $\mathbf{\hat{s}}[n]$, HuBERT's output $\mathbf{\hat{S}}_\mathrm{OL}$ is also the best performing achieving near identical performance to the $\mathbf{\hat{s}}'[n]$ model. In terms of the difference in performance between using earlier \ac{SSSR} representations $\mathcal{G}_\mathrm{FE}$ or output representations $\mathcal{G}_\mathrm{OL}$ as features, this seems to depend on the \ac{SSSR} used; for HuBERT the output layers perform best, while for XLSR the feature encoder layers show better performance.   
\begin{table}[!t]
\caption{Non Intrusive Prediction Performance on the CPC1 open set.}
\label{tab:open-results}
\begin{tabular}{l|cccc}
\textbf{Model Name} & \textbf{RMSE} & \textbf{Var} & \textbf{Spearman} & \textbf{Pearson} \\ \hline
\textit{CPC 1 Baseline}            & \textit{36.50}         & \textit{--}           & \textit{0.53}              & \textit{--}               \\
\textit{E23}~\cite{McKinney_2022}                 & \textit{43.70}         & \textit{--}           & \textit{0.05}              & \textit{-- }              \\
%E35~\cite{McKinney_2022}                 & 35.70         & --           & 0.22              &                  \\ 
\textit{E33}~\cite{edozezario22_interspeech}                  &\textit{ 28.9}                           &\textit{--}                            & \textit{0.65}
& \textit{--}                                   \\ 
\textit{E29}~\cite{tu22b_interspeech}                 & \textbf{\textit{24.60}}         & \textit{--}           & \textbf{\textit{0.73}}              & \textit{--   }            \\\hline
$\mathrm{SPEC}~\mathbf{\hat{S}}_\mathrm{SPEC}$           &  32.84&	1.29&	0.35&	0.50           \\
$\mathrm{SPEC}~\mathbf{\hat{S}'}_\mathrm{SPEC}$               & 29.16         & 1.15         & 0.57              & 0.60             \\
$\mathrm{HuBERT}~ \mathbf{\hat{S}}_\mathrm{FE}$&  33.69&	1.30 &	0.27&	0.45        \\
$\mathrm{HuBERT}~ \mathbf{\hat{S}'}_\mathrm{FE}$      & 35.31         & 1.40         & 0.19              & 0.24             \\
$\mathrm{HuBERT}~ \mathbf{\hat{S}}_\mathrm{OL}$          & 32.43 &	1.22 &	0.47 &	0.54             \\
$\mathrm{HuBERT}~ \mathbf{\hat{S}'}_\mathrm{OL}$          & 29.66         & 1.14         & 0.60              & 0.61             \\
$\mathrm{XLSR}~  \mathbf{\hat{S}}_\mathrm{FE}$         & 31.83 &	1.26 &	0.49 &	0.52             \\
$\mathrm{XLSR}~  \mathbf{\hat{S}'}_\mathrm{FE}$         & 30.86         & 1.19         & 0.56              & 0.56             \\
$\mathrm{XLSR}~  \mathbf{\hat{S}'}_\mathrm{OL}$         &  31.85 &	1.25 &	0.42 &	0.49 \\
$\mathrm{XLSR}~  \mathbf{\hat{S}'}_\mathrm{OL}$            & 34.54         & 1.36         & 0.26              & 0.37      
\end{tabular}
\end{table}

\subsection{Results of CPC1 Open set}
\autoref{tab:open-results} shows the performance of the proposed systems on the more challenging CPC1 open set (cf.~\autoref{sec:dataset}). Performance of the proposed systems is significantly worse than that of the closed set for all systems, with a much larger variance in MSE in all cases, but all proposed systems still outperform the baseline.  The poorer performance might be due to overfitting of the models to the training data, (in particular to the enhancement systems in the training set) as the test data contains unseen enhancement systems and listeners. All of the models here perform similarly poorly.  
%\vspace*{-1ex}
\subsection{System and Listener-wise Analysis}
\begin{figure}[!ht]
 \centering
 \resizebox{1\columnwidth}{!}{%
 \def\svgwidth{\columnwidth} 
 	\graphicspath{{figs/}} 
 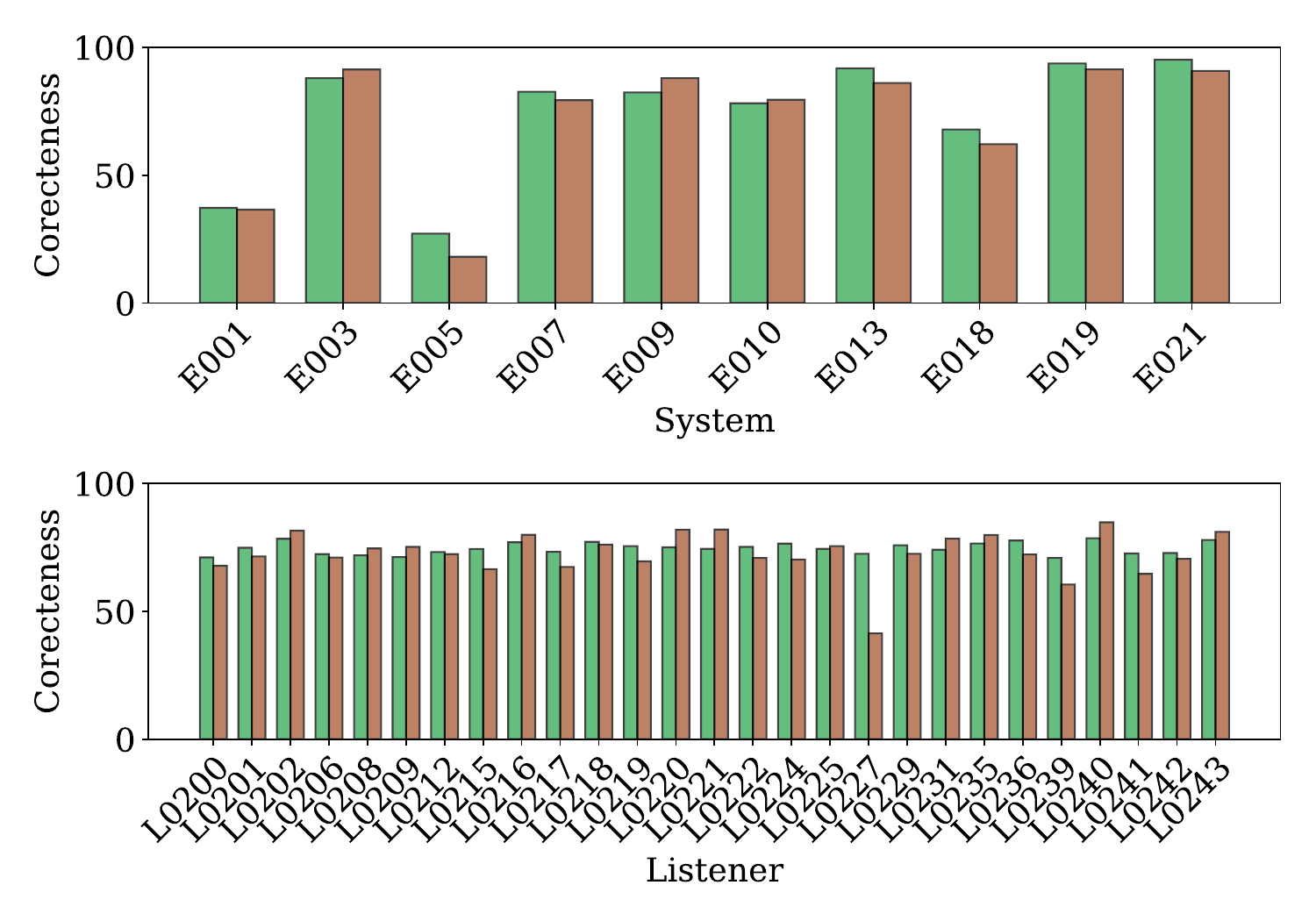 
 }
\caption{System (top) and listener-wise (bottom) correctness prediction $\hat{i}$ (l./green) vs.~true $i$ (r./brown) using HuBERT output for $\mathbf{\hat{S}}_\mathrm{OL}$ model on CPC1 closed set. }\label{fig:no_hl_detail_plots}
\end{figure}

\begin{figure}[!ht]
 \centering
 \resizebox{1\columnwidth}{!}{%
 \def\svgwidth{\columnwidth} 
 	\graphicspath{{figs/}} 
 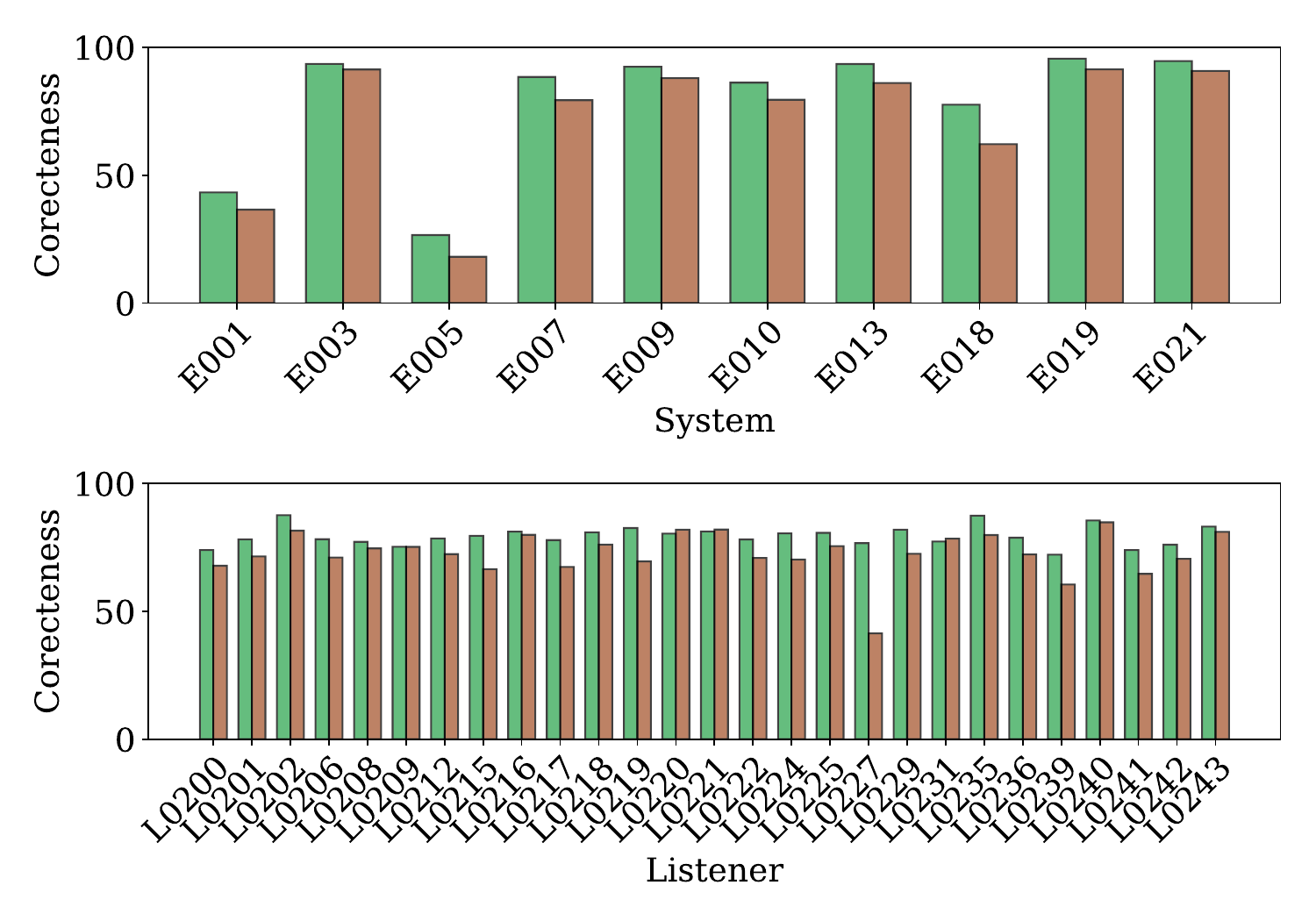 
 }
\caption{System (top) and listener-wise (bottom) correctness prediction $\hat{i}$ (l./green) vs.~true $i$ (r./brown) using HuBERT outpur for $\mathbf{\hat{S}'}_\mathrm{OL}$ model on  CPC1 closed set.}\label{fig:hl_detail_plots}
\end{figure}
%%%
%%%
\begin{figure}[!ht]
 \centering
 \resizebox{1\columnwidth}{!}{%
 \def\svgwidth{\columnwidth} 
 	\graphicspath{{figs/}} 
 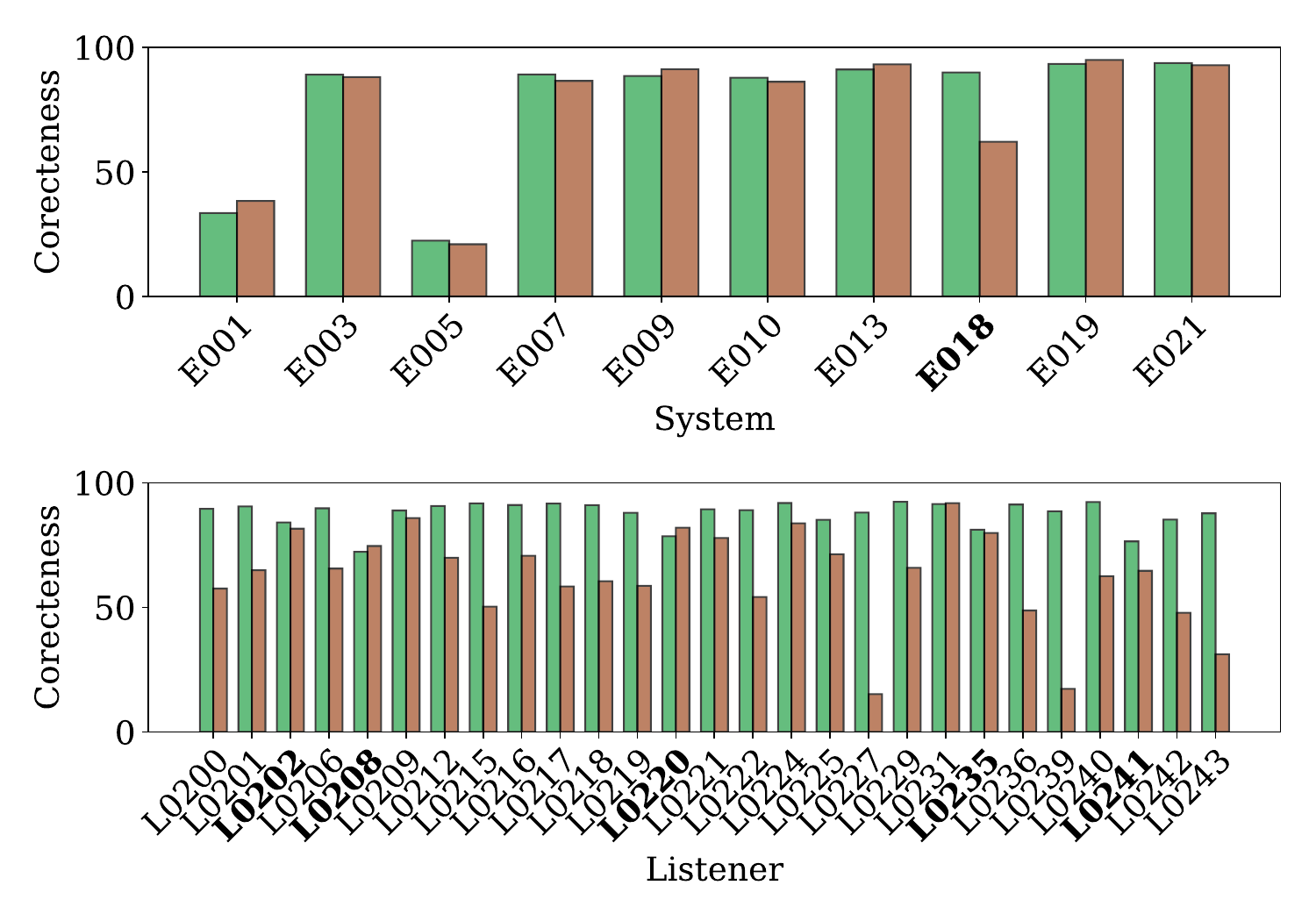 
 }
\caption{System (top) and listener-wise (bottom) correctness prediction $\hat{i}$ (l./green) vs.~true $i$ (r./brown) using HuBERT output for $\mathbf{\hat{S}}_\mathrm{OL}$ model on  CPC1 open set. Listeners and Systems unseen during training are bold.}\label{fig:hl_detail_plots_indep}
\end{figure}
%%%%%%%
\begin{figure}[!ht]
 \centering
 \resizebox{1\columnwidth}{!}{%
 \def\svgwidth{\columnwidth} 
 	\graphicspath{{figs/}} 
 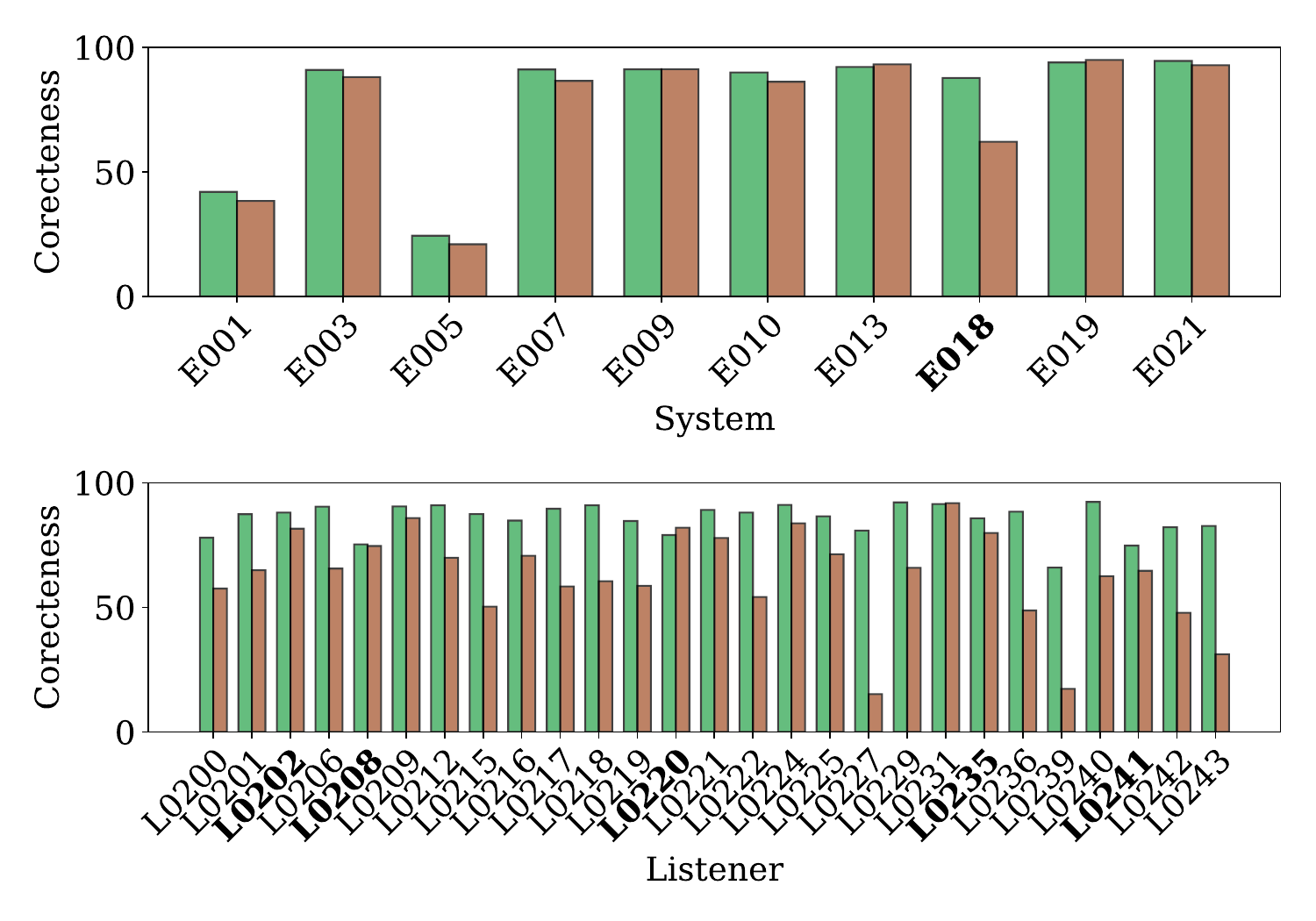 
 }
\caption{System (top) and listener-wise (bottom) correctness prediction $\hat{i}$ (l./green) vs.~true $i$ (r./brown) using HuBERT output for $\mathbf{\hat{S}'}_\mathrm{OL}$ model on  CPC1 open set. Listeners and Systems unseen during training are bold.}\label{fig:no_hl_detail_plots_indep}
\end{figure}
%%%%%%%%%%%%%%%%%%%%%%%%%%%%%%%%%%%%%%
\noindent For further analysis, Figs.~\ref{fig:no_hl_detail_plots} and \ref{fig:hl_detail_plots} show ground truth and predicted correctness across the hearing aid systems and across the listeners in the CPC1 open testset for the HuBERT $\mathbf{\hat{S}}_\mathrm{OL}$ and HuBERT $\mathbf{\hat{S}'}_\mathrm{OL}$ models, respectively. Both models show similar performance across the different hearing aid systems, both successfully assigning low scores to the audio enhanced by the E005 hearing aid system. This indicates that the models are able  (at some level) to detect the distortions introduced by this enhancement. Similarly, there is little difference in performance across the subset of listeners for the two models; this suggests that the listener-specific hearing loss information which the $\mathbf{\hat{S}'}_\mathrm{OL}$ model has access to (encoded in the audio) does not aid in the intelligibility prediction performance. It should be noted that already the enhancement system (hearing aid) has (implicitly) access to the hearing loss information and is expected to process its input signal accordingly (cf.~\autoref{fig:Clarity_schematic}). Interestingly, both models overestimate the intelligibility ratings of speaker L0227 who performs worse than average at the intelligibility task (cf.~\autoref{fig:intel_bar}). This suggests that L0227's lower performance is not due to their hearing loss but rather other unknown factor(s); audiogram information for this listener does not show that they have particularly severe hearing loss.\\
Figs.~\ref{fig:no_hl_detail_plots_indep} and \ref{fig:hl_detail_plots_indep} show ground truth and predicted correctness across the hearing aid systems and across the listeners for the more challenging CPC1 closed testset for the HuBERT output for $\mathbf{\hat{S}}_\mathrm{OL}$ and HuBERT output for $\mathbf{\hat{S}'}_\mathrm{OL}$ models, respectively. Systems and listeners which are unseen during the training of the models are highlighted by bold-font. Here, the overfitting of the proposed system to the hearing aid systems during training can be observed by the poor performance on the unseen hearing aid system in the testset, E018. The overall lower performance of the proposed systems on the closed set is shown by the listener-wise plots, with both systems significantly overestimating the correctness versus the true value; however the encoding of the hearing loss information in  $\mathbf{\hat{S}'}_\mathrm{OL}$ does appear to have some positive effect here.

\section{Conclusion}
This work explores the use of \ac{SSSR} models as feature extraction for non-intrusive speech intelligibility prediction networks in comparison to traditional, spectrogram-based input. Both, the final \ac{SSSR} representation and the intermediate output of the \ac{SSSR} feature encoder are compared for the first time for an \ac{SI} prediction task for hearing-impaired users. Results indicate that encoding the hearing loss of a particular listener via (an additional) hearing loss simulation does not typically improve performance. Additionally, models tend to overfit to specific hearing aid systems, as demonstrated by the results on the open set which might be alleviated by larger datasets released in the future. The upcoming CPC2 challenge data, which includes twice the number of enhancement systems, could mitigate this issue. %Ultimately, there are unidentified factors that substantially impact intelligibility but are not captured.

\bibliographystyle{IEEEbib}
\bibliography{strings,refs}

\end{document}